\documentstyle[12pt,aasms]{article}

\begin{document}

\title{Effect of Screening on Thermonuclear Fusion in Stellar and
Laboratory Plasmas}

\author{L. Wilets$^a$, B. G. Giraud$^b$, M. J. Watrous$^a$ and J. J. Rehr$^a$}
\smallskip
%\address{$^a$Department of Physics, Box 351560, 
\affil{$^a$Department of Physics, BOX 351560, Univ. of Washington,
 Seattle, WA 98195-1560, USA 
\noindent$^b$Service de Physique Th\'eorique, CE Saclay, F-91191
    Gif-sur-Yvette, France}

%\date{\today}
%\maketitle

\begin{abstract}
The fusion enhancement factor due to screening in the solar plasma is
calculated. We use the finite temperature Green's function method and
a self consistent mean field approximation. We reduce this to one 
center problems, because in the collision of two fusing ions, the
turning point where tunneling may occur lies far inside the screening 
radius. The numerical results given by this method indicate that screening 
may be slightly weaker than that obtained in the most recent previous 
calculations. 

\end{abstract}
 
\section{Introduction}

There remains continued interest in the solar neutrino problem which
has not been resolved in terms of the standard solar model and standard
particle physics.  Neutrinos are generated in thermonuclear reactions, and
thermonuclear reaction rates are determined by solar physics (temperature 
and density), nuclear physics, and atomic shielding.
 The solar neutrino problem is 
just one of several examples where there is current interest in thermonuclear
reaction rates; others include general stellar structure and laboratory plasmas.
 We address here
the role of atomic shielding, using a finite temperature Green's function
method.  Recent calculations of screening effects in the sun
have been reported by 
%Gruzinov and Bahcall
\cite{gb},
who also review previous solar screening calculations.
Results of those calculations, with which we compare our present work,
exhibit a scatter of 10 to 30\% or more in the effect of
screening on fusion rates.
As discussed in \cite{gb}, all of these previous 
calculations are based either on Salpeter's weak screening formula, which
is a classical approximation, or else make 
unphysical approximations in order to go beyond the linear regime.  
Gruzinov and Bahcall attempted to rectify these
limitations by including quantum effects, treating the kinetic energy in 
perturbation theory.  Out model differs from theirs in that we do not rely 
on perturbation theory.

The physical model we employ is based on self-consistent
Hartree plus finite temperature, local density approximation
for exchange and correlation and assumes
thermal equilibrium and adiabaticity in the 
internuclear coordinates.  We solve this model making no further 
approximations or expansions, except for numerics which are carefully 
monitored.  All previous calculations are essentially approximations to 
this model, involving {\it e.g.} expansions in the temperature, density or 
quantum corrections. \cite{csk}
%Carraro, Sch\"afer and Koonin\protect\cite{csk}
also consider dynamic effects.
 Thus we believe that our approach yields results which are the most accurate 
to date. These results yield fusion enhancement factors which are typically
 a little smaller than those recently reported (\cite{gb}).
%by Gruzinov and Bahcall \cite{gb}.

The rate of thermonuclear fusion in plasmas is governed by barrier
penetration. The barrier itself is dominated by the Coulomb repulsion of the
fusing nuclei. Because the barrier potential appears  in the exponent of
the Gamow formula, the result is very sensitive to the effects of screening
by electrons and positive ions in the plasma.  Screening lowers the barrier
and thus enhances the fusion rate; it is more important the greater the 
nuclear charges, and thus plays an important role in the solar 
neutrino spectrum.

As shown below, the classical turning point radius which enters in the WKB
integral for barrier penetration is very small compared with the
characteristic screening lengths of interest.  Inside this radius, the
barrier potential is just the nucleus-nucleus Coulomb repulsion minus a
constant due to screening.  The constant can be interpreted simply as
the difference in free energies of the system for either united and separated
nuclei. Since both of these are spherically symmetric, one needs only consider
one-center problems. 

We present here numerical calculations relevant to the solar core.  
The screening due to electrons in the plasma is calculated quantum 
mechanically by a novel Green's function method described in a recent paper 
%by Watrous, Wilets and Rehr
(\cite{wwr}). The screening due to the ions is 
treated classically but self-consistently with the electrons.  The resultant 
enhancement factors for several nuclear reactions are presented and compared 
with earlier results by other authors.

\section{Fusion rate}

       We work in atomic units, $e=\hbar=m_e=k_B=1;$ the unit of atomic 
temperature is $3.159\times 10^5 \, K$. 
       As shown by
%Salpeter
\cite{sal} (here we follow
%Clayton
\cite{clay}), 
the fusion reaction rate between species 1 and 2 is given by
\begin{equation}
r_{1,2}=N_1 N_2\int_0^\infty \psi(E)v(E)\sigma(E)\,dE  ,
\end{equation}       
where $N_1$ and $N_2$ are the number densities of the colliding nuclides, 
$\psi(E)$ is the Maxwellian probability that the center-of mass-energy at 
large separation is $E$; and the cross section $\sigma$ is written as a product 
of a penetrability factor $P$ and a nuclear factor $\sigma_{nuc},$
\begin{equation}
\sigma(E)=P(E)\sigma_{nuc}(E)\,.
\end{equation}
In the WKB approximation, the s-wave penetration is given by
\begin{equation}
P_0(E)=\left({E_B\over E}\right)^{1/2}\exp\left[-2\sqrt{2\mu}
\int_R^{R_0}\big[Z_1Z_2/r+U_{sc}(r)-E\big]^{1/2}dr\right]\,,
\end{equation}
with $E_B$ the height of the barrier, $\mu$ the reduced mass and 
$U_{sc}(r)<0$ the screening potential; $R$ is the ``touching" radius 
(the top of the barrier), and $R_0$ is the classical turning radius.
The integrand in Eq.~(1) peaks at some characteristic energy $E_0$.  For the 
solar interior, the corresponding classical turning radius
$$R_0\approx Z_1 Z_2/E_0,$$
is $\sim 0.005$ au compared with a screening radius of $\sim 0.5$ au.
For this reason, the 
screening potential can be taken as a constant $U_0\equiv U_{sc}(r=0)<0$ 
within the turning radius, acting as an effective shift in the energy:
\begin{eqnarray}
r_{1,2}&=&N_1 N_2\int_0^\infty \psi(E) v(E)P(E-U_0)\sigma_{nuc}(E-U_0)\,dE\\
       &\propto&\int_0^\infty E^{1/2}e^{-E/T}P(E-U_0)\sigma_{nuc}(E-U_0)\,dE\\
       &=&\int_{-U_0}^\infty (E'+U_0)^{1/2}e^{-(E'+U_0)/T}P(E')\sigma_{nuc}
        (E')\,dE'\,.
\end{eqnarray}       
Then, neglecting $U_0$ in the limit of integration,
\begin{equation}
r_{1,2}\approx e^{-U_0/T}N_1 N_2\int_0^\infty \psi(E)v(E)\sigma(E)\,dE\,.
\end{equation}
Screening thus enhances the reaction rate by the factor
\begin{equation}
  f=e^{-U_0/T}\,,
\end{equation}
where $U_0$ is given by Eq. (10) below.
Since $U_0<0$, $f>1$, corresponding to enhancement.
\section{Free Energy and the One Center Problem}

      The screening potential is just the change, brought about
by the approaching ions, in the Helmholtz free energy
\begin{equation}
F=U_{internal}-T\,S ,
\label{eq:f1}
\end{equation}
as a function of nuclear separation.  Because of the smallness 
of the turning radius, it is sufficient to consider the one-center problems 
and identify
\begin{equation}
U_0=F(Z_1+Z_2)-F(Z_1)-F(Z_2)\,.
\end{equation}
What we mean here by one-center problems are the calculations of $F$ 
when either
charges $Z_1,Z_2$, or both lie at the center of the plasma.
In Sec. IV we calculate $F(Z)$ explicitly for an independent particle model 
approximation, akin to the temperature dependent Hartree-Fock method.
However, the free energy has a simple and completely general
(for the problem at hand)
interpretation, which does not depend on the mean field approximation.  It is,
by definition, the work required at a given temperature to introduce the
nuclear charge into the plasma.  The nucleus only interacts with the plasma
electrostatically, so $dF=\phi(Z',r=0)dZ'$, and thus the work required to
assemble the charge $Z$ at, say, $r=0$ is
\begin{equation}
F(Z)=\int_0^Z \phi(Z',r=0)\,dZ'\,,
\label{eq:f2}
\end{equation}
where $\phi(Z',r)$ is the electrostatic potential generated in the plasma 
due to a nuclear charge $Z'$,
\begin{equation}
\phi(Z', r=0)=4\pi\int_0^\infty r'dr'\Big[\sum_I 
\rho_I(r',T)-\rho_e(r',T)\Big]\,.
\label{eq:phi1}
\end{equation}
Here $\rho_e(r,T)$ and $\rho_I(r,T)$ denote the finite temperature number
density (recall $e=1$)
of the electrons and various background ions charge densities, respectively.

      The one-center screening problem is solved by the methods described by
\cite{wwr} for the electrons with a Kohn-Sham formalism, using the finite 
temperature local density approximation (\cite{Mermin65}) for exchange and 
correlation.  Screening due to ions is included as well as that due to 
electrons.  The ions are treated classically according to the Debye-Huckel 
method.  Thus the electric potential generated by electrons and ions becomes
\begin{equation}
\Phi( r)={Z\over r}+\int d^3r'{\sum_I \rho_I(r',T)-\rho_e(r',T)\over 
       |{\bf r-r'}|}\,.
\end{equation}
The second term is the function denoted by $\phi(Z,r)$ above. 
Each ion density is given by
\begin{equation}
\rho_I(r,T)=\rho_I(\infty)e^{-Z_I\Phi(r)/T}\,.
\end{equation}
Since $\Phi(r)$ is positive, ions are pushed outward and electrons are 
drawn toward the nucleus. For notational simplicity in the following, 
the temperature dependence of the densities will be understood and
not specified explicitly.

\section{Alternative Mean Field Derivation of the Free Energy}

       We now demonstrate the equivalence of the $F$ defined by equations
(\ref{eq:f1}) and (\ref{eq:f2}) in the
finite-temperature Hartree(-Fock) approximation, which is similar to
the finite temperature local density approximation
actually used in our numerical calculations.
In a transparent, short notation, we describe the system 
by a second-quantized Hamiltonian ${\cal H}$ containing the kinetic energy 
operators $t_i$ of the particles present in the plasma, their two-body 
interactions $v_{ij},$ and the additional one-body operators $w_i=-Z/r_i$
or $+Z_I\,Z/r_i$
representing the contributions of the additional nuclear charge $Z$ at the 
center. The temperature $T$ enters the formalism via the 
the usual Boltzman factor $\exp(-\beta {\cal H})$ where $\beta=1/T$. Actually, 
because the average electronic density and background positive charge density
are fixed parameters of the problem, ${\cal H}$ must be replaced by the usual
constrained Hamiltonian. This will read, in a short notation, 
${\cal H}-\mu'{\cal N}.$ Here $\mu'$ is a several-component chemical 
potential because of the several components in the plasma (i.e., electrons
and various background ions).
Accordingly the particle number operator ${\cal N}$ must be understood as a 
several-component operator.
     For simplicity, however, the following equations deal with 
the electrons only. This is because the positive ions in the 
background may consist of bosons as well as fermions,
and we want to spare the reader the cumbersome symmetrization or 
antisymmetrization formulae for such backgrounds. In any case, 
the present paper describes the background density classically, and
this density is not high enough to demand exchange terms.

     The independent particle ansatz approximates the eigenstates of 
${\cal H}-\mu' {\cal N}$ by antisymmetrized products of single 
particle states (orbitals) $|i\rangle,$ with an approximate spectrum made of 
sums of independent particle levels $\eta_i.$ The infinite set of orbitals 
$|i\rangle$ and eigenvalues $\eta_i$ define a one-body operator 
${\cal H}_0-\mu' {\cal N}$, presumed optimal from the point of view of 
a variational principle: i.e., stationarity of the grand potential 
or free energy $F$.

     In second quantization, the many-body density matrix used as a
{\it trial} density operator (noted in the following by
quantities carrying a subscript $t$) is
\begin{equation}
D_t={\exp[-\beta({\cal H}_0-\mu' {\cal N})]\over Z_t}=
{\exp(-\beta\sum_i\eta_ic^{\dagger}_ic_i)\over Z_t},
\end{equation}
where $c^{\dagger}_i$ and $c_i$ are the usual fermionic creation and 
annihilihation operators for orbital $|i\rangle$. For electrons
the normalization denominator $Z_t$ is 
\begin{equation}
Z_t=\prod_i \left[\ 1+\exp(-\beta\eta_i)\ \right],
\end{equation}
so the trial density becomes
\begin{equation}
D_t=\prod_i
{\exp(-\beta \eta_i c^{\dagger}_ic_i) \over 1+\exp(-\beta \eta_i) }.
\end{equation}
   The {\it true} density matrix
$D= { \exp[-\beta({\cal H}-\mu' {\cal N})] /
{\rm Tr}\, \exp[-\beta({\cal H}-\mu' {\cal N})] }$,
minimizes the {\it true} grand potential,
$\Omega= {\rm Tr}\ D ({\cal H}-\mu' {\cal N}+\beta^{-1}\ln D)$.
However, with an independent particle  approximation one must be
content with minimizing
\begin{equation}
F= {\rm Tr}\ D_t ({\cal H}-\mu' {\cal N}+\beta^{-1}\ln D_t).
\end{equation}
It is easy to show that
\begin{equation}
{\rm Tr}\ D_t c^{\dagger}_ic_i c^{\dagger}_jc_j=f_i f_j,
\ \ {\rm if}\ \ i\ne j, 
\end{equation}
where the Fermi occupation numbers are 
$f_i\equiv {\rm Tr}\ D_t c^{\dagger}_ic_i = 1/[1+\exp(\beta \eta_i)]$.
Furthermore
\begin{equation}
{\rm Tr}\ D_t
\ \{-\beta \eta_i c^{\dagger}_i c_i - \ln [1+\exp(-\beta\eta_i)]\}=
f_i \ln f_i+(1-f_i)\ln (1-f_i).
\end{equation}
Finally, because ${\cal H}=({\cal T}+{\cal W})+{\cal V}$ is the sum of a 
one-body operator ${\cal T}+{\cal W} \equiv \sum_i t_i+\sum_i w_i$ and 
a two-body operator ${\cal V}\equiv \sum_{i>j}v_{ij},$ we obtain
\begin{equation}
F=
\sum_i f_i \langle i|(t+w-\mu')|i\rangle +{1\over 2}\sum_{i,j} f_i f_j
\langle ij|\tilde v|ij\rangle +
\beta^{-1}\sum_i [f_i \ln f_i+(1-f_i)\ln (1-f_i)].
\end{equation}
Here the tilde $\tilde v$ means that the matrix element
$\langle ij|\tilde v|ij\rangle$ is antisymmetrized. 

     The functional derivative of $F$ with respect to $\langle i|$ is then
\begin{equation}
{\delta F \over \delta \langle i|}= 
f_i\ \left(t+w-\mu'+\sum_j f_j \langle .j|\tilde v|ij\rangle\right),
\end{equation}
where one recognizes the action of a mean field potential,
including both direct and exchange terms
\begin{equation}
{\cal U}|i\rangle \equiv \sum_j f_j \langle .j|\tilde v|ij\rangle.
\end{equation}
Stationarity of $F$ with respect to $|i\rangle$ then gives the ``finite 
temperature Hartree-Fock'' equations,
\begin{equation}
(t+w-\mu'+{\cal U})|i\rangle=\eta_i|i\rangle,\ \ \ \forall\, i,
\end{equation}
where one recognizes that $\eta_i$ is the Lagrange multiplier for
the normalization of the orbital.

     In the same way, the derivative of $F$ with respect to $\eta_i,$
or as well $f_i,$ yields
\begin{equation}
{\partial F \over \partial f_i} = \eta_i+\beta^{-1}\ln \left[
{f_i/1-f_i}\right],
\end{equation}
which vanishes [see Eq.~(20)].  It can thus be concluded that 
\begin{equation}
F=-\beta^{-1}\ln Z_t-{\rm Tr}D_t{\cal V},
\end{equation}
(which {\it differs} from $-\beta^{-1}\ln Z_t$) is stationary with respect
to variations of $D_t.$ Therefore, since ${\cal W}$ 
is proportional to the additional charge $Z$ at the center, 
the derivative of $F$ with respect to $Z$ must be given by
\begin{equation}
{ \partial F \over \partial Z} = {\rm Tr}\ D_t \, { {\cal W} \over Z}.
\end{equation}
\label{eq:f3}
Notice also that ${\cal W}$ is a local potential. Hence,
in the coordinate representation, only the diagonal part of $D_t$ 
(the classical density) is needed to calculate $\partial F / \partial Z,$ 
\begin{equation}
{\partial F \over \partial Z}=
4 \pi \int r' d r'\,\Big[\rho_p(r')-\rho_e(r')\Big],
\end{equation}
which is the differential form of Eqs.~(\ref{eq:f2}-\ref{eq:phi1}).
Here $\rho_e(r')$ and $\rho_p(r')$ are
diagonal matrix elements $\langle r'|D_t|r'\rangle$ 
in the electron and positive background sectors, respectively. It will
be noticed that Eq.~(28) gives the background density as well as the 
electronic density, while the preceding equations, Eqs.~(15-27), accounted 
for the electrons only. In view of the simplicity of the result due to the 
electrons, this reinstatement of the background contribution is trivial. 
Notice also that for the one-center problem, $D_t$ is rotationally 
invariant, hence no vector label ${\vec r}\,'$ is needed.

     The result, Eq.~(28) is simply a reformulation of Eqs.~(11-12). 
The method of Matsubara poles used by \cite{wwr}
%in Ref.\cite{wwr},
via the finite temperature one-body Green's function, is perfectly suited to 
this local calculation of the density and avoids an explicit solution of
Eqs.~(24). Indeed, as discussed in detail by \cite{wwr},
%in Ref.\cite{wwr},
a local density
such as $\rho_e(r')$ can be directly derived from diagonal matrix elements
$\langle r'|(\eta-t-w-\mu'-{\cal U})^{-1}|r'\rangle$ of the one-body Green's
function.  Such matrix elements are integrated on a suitable contour in the 
$\eta$-complex plane.
     Once $\partial F / \partial Z$ is known for all values of $Z$ smaller 
than $Z_1+Z_2,$ trivial integrals provide the screening $U_0$ according to 
Eq.~(10).

\section{NUMERICAL RESULTS FOR THE SOLAR CORE}

       Shielding calculations have been performed for $Z=1, 2, 4, 6, 8$ 
at a density and temperature relevant to energy and neutrino production in 
the sun.  From these calculations the fusion enhancement factors are
calculated and compared with the results of other researchers.
%       Ricci, {\it et al.}
\cite{ricci} have shown that enhancement factors 
for the relevant fusion reactions are insensitive to the location within the 
solar core.  We use here the set of parameters similar to those employed by 
%Bahcall and Pinsonneault
\cite{b&p}, corresponding to $R/R_\odot=0.06$.  
In atomic units, 
they are $T= 47,\quad n_e= 7.63,\quad X= 0.432,\quad Y=0.568.$
       The results are displayed in Table 1 for $\phi(Z, r=0)$ and 
$\Delta\rho_e(Z,r=0)=\rho_e(Z,r=0)-n_e$.  
These quantities for other $Z$-values can be obtained by 
interpolation.  $\phi(Z, r=0)$ was fitted to a fourth order polynomial,
\begin{equation}
\phi(Z, r=0)\approx-\sum_{n=1}^4c_n\,Z^n.
\end{equation}
The integral of this quantity then yields the free energy,
\begin{equation}
F(Z)\approx-\sum_{n=1}^4c_n\,Z^{n+1}/n\,,
\label{eq:f4}
\end{equation}
with $c_1=2.0275,\ c_2=-0.03661,\ c_3=0.00594,\ c_4=-0.000103.$
Note that $\phi(Z,r=0)$ is nearly linear in $Z$.

Table II gives the fusion enhancement factors for several reactions of 
interest in the solar neutrino problem.  
The deviation from unity is due to shielding.  Note that the 
various calculations exhibit a scatter of 10 to 30\% and more in the deviation 
from unity.  Although our factors are rather close to the recent
calculations of
%Gruzinov and Bahcall
\cite{gb}, they are usually somewhat lower.

\section{Summary and Conclusions}
The finite temperature Green's function method of \cite{wwr}
has been applied to the problem of screening of the nucleus-nucleus interaction
in the solar plasma. The method is based on a self consistent,
finite temperature Kohn-Sham formalism with the local density
approximation for exchange and correlation.  Atomic bound states are
included on an equal footing with continuum states.
Fusion enhancement factors (over pure Coulomb) are calculated for various 
relevant nuclear reactions at mean conditions (temperature and density) 
of the solar core.  Comparisons with several other calculations are presented.

The method appears to have no restriction with respect to temperature or
density for stars in ``non catastrophic,'' thermal equilibrium  states. 
When implementing 
calculations for other systems, one may need to extend the range of 
parameters for the local exchange-correlation term, which is quite small 
for our solar system.

\section{Acknowledgments}

This work was supported in part by the U. S. Department of Energy
and by the U. S. National Science Foundation. 

\newpage

\begin{table}[t]
        \begin{center}
        \caption{Results of solar screening calculations for various values
of $Z$.  $\phi$ and $\Delta\rho_e$ are calculated values; the free energy F is 
based on the polynomial fit.} 
        \begin{tabular}{cccr}
                $Z$   & $\phi(Z,r=0)$         
                        & $\Delta\rho_e(Z,r=0)$      
                        & \multicolumn{1}{c}{$F(Z)$}
                                                                        \\
                \hline
                1 & 2.007  & 2.791 & -1.008 \\
                2 & 3.944  & 6.829 & -4.009 \\
                3 & --- & ---      & -8.962 \\
                4 & 7.884  & 20.76 & -15.82 \\ 
                5 & ---    & ---   & -24.54 \\ 
                6 & 12.00  & 49.77 & -35.14 \\ 
                7 & ---    & ---   & -47.65 \\ 
                8 & 16.51  & 108.6 & -62.19 \\ 
        \end{tabular}
        \label{Table:1}
        \end{center}
\end{table}
\bigskip
\bigskip
\bigskip

\begin{table}
        \begin{center}
        \caption{Fusion enhancement factors for the solar interior (This 
Work), compared 
with other screening calculations: WES is Salpeter's weak screening 
approximation\protect (\cite{sal}); Mit is Mitler's Thomas-Fermi-like 
model\protect (\cite{mit});
GDGC is due to
%Graboske {\it et al.}
\protect\cite{grab}; CSK is the dynamic screening model of
%Carraro, Sch\"afer and Koonin
\protect\cite{csk}; 
GB is due to
%Gruzinov and Bahcall
\protect\cite{gb}.} 
        \begin{tabular}{ccccccr}
                reaction  & This Work & WES         
                        & Mit & GDGC & CSK
                        & \multicolumn{1}{c}{GB}
                                                                        \\
                \hline
                $p+p$ & 1.043 & 1.049 & 1.045 & 1.049 & 1.038 & 1.053 \\
                He+He & 1.181 & 1.213 & 1.176 & 1.115 & 1.158 & 1.224 \\
                Be+$p$& 1.183 & 1.213 & 1.171 & 1.112 & 1.169 & 1.166 \\
                N+$p$ & 1.356 & 1.402 & 1.293 & 1.191 & 1.324 & 1.393 \\ 
        \end{tabular}
        \label{Table:2}
        \end{center}
\end{table}

%
%************************************************************************** 

%\begin{thebibliography}{99}

\end{document}